\documentclass[prb,twocolumn]{revtex4}%

\newcommand{\be}{\begin{equation}}
\newcommand{\ee}{\end{equation}}
\newcommand{\bea}{\begin{eqnarray*}}
\newcommand{\eea}{\end{eqnarray*}}
\newcommand{\bean}{\begin{eqnarray}}
\newcommand{\eean}{\end{eqnarray}}

\begin{document}

\draft
\title
{\bf Bipolar thermoelectric effect in a serially coupled quantum dot
system}

\author{ David M.-T. Kuo$^{1\dagger}$, and Yia-chung Chang$^{2*}$  }
\address{$^{1}$Department of Electrical Engineering and Department of Physics, National Central
University, Chungli, 320 Taiwan}

\address{$^{2}$Research Center for Applied Sciences, Academic
Sinica, Taipei, 115 Taiwan}

\date{\today}

\begin{abstract}
The Seebeck coefficient (S) of a serially coupled quantum dot (SCQD)
junction system is theoretically studied via a two-level Anderson
model. A change of sign in S with respect to temperature is found,
which arises from the competition between tunneling currents due to
electrons and holes (i.e, bipolar tunneling effect). The change of
sign in S implies that one can vary the equilibrium temperature to
produce thermoelectric current in either the forward or reverse
direction, leading to a bipolar thermoelectric effect. For the case
of two parallel SCQDs, we also observe the oscillatory behavior of S
with respect to temperature.
\end{abstract}

\maketitle



 Owing to energy and environment issues, it has become important to
consider novel applications related to the thermal properties of
materials. Many considerable studies have been devoted to seeking
efficient thermoelectric materials because there exist potential
applications of solid state thermal devices such as coolers and
power generators.$^{1-9)}$ A quantum dot-(QD) based thermal device
was also predicted to have more pronounced enhancement in energy
conversion.$^{9)}$ Recently, some theoretical efforts have focused
on the thermoelectric effects in nanostructure junctions,$^{10-13)}$
however, not many works have paid attention to the thermoelectric
effects of a serially coupled quantum dot (SCQD) junction, which
exhibits features of current rectification due to spin blockade,
negative differential conductance, nonthermal broadening of
electrical conductance, and coherent tunneling (for identical QDs)
in the Coulomb blockade regime.$^{14)}$ Our recent work has
described these observed phenomena in a unified theory.$^{15)}$
Based on our previous work, we find that the Seebeck coefficient of
SCQDs exhibits a behavior of sign change with respect to temperature
arising from electron Coulomb interactions.


Using the Keldysh-Green's function technique,$^{15}$ we can express
(up to the second order in the interdot coupling, $t_c$) the
tunneling current through a serially coupled QDs connected to
metallic electrodes (shown in the inset of Fig. 1)
\begin{eqnarray}
J&=&\frac{2e}{h}\int d\epsilon {\cal T}(\epsilon)
[f_L(\epsilon)-f_R(\epsilon)]
\end{eqnarray}
where ${\cal T}(\epsilon)\equiv \Gamma_L(\epsilon)
\Gamma_R(\epsilon) ({\cal A}_{12} +{\cal A}_{21})/2$ is the
transmission factor. $\Gamma_{\ell=L,R}(\epsilon)$ denote the tunnel
rate from the left electrode to dot A and the right electrode to dot
B. $f_{L(R)}(\epsilon)=1/[e^{(\epsilon-\mu_{L(R)})/k_BT_{L(R)}}+1]$
denotes the Fermi distribution function for the left (right)
electrode. The chemical potential difference between these two
electrodes is related to $\mu_{L}-\mu_{R}=e \Delta V$. $T_{L(R)}$
denotes the equilibrium temperature of the left (right) electrode.
$e$ and $h$ denote the electron charge and Plank's constant,
respectively. For simplicity, we consider wide-band limit that is
$\Gamma_{\ell}(\epsilon)=\Gamma_{\ell}$. ${\cal A}_{\ell,j}$ denotes
the spectral density, which can be calculated by one particle off
diagonal Green's function.$^{15}$ In the atomic limit, we have

\begin{equation}
{\cal A}_{\ell,j}(\epsilon)=t^2_{c}\sum_m p_m/|\Pi_m|^2 \label{3}; \; (\ell \ne j),
\end{equation}
where the numerators denote probability factors for various charge
configurations, and they are
$p_1=(1-N_{\ell,\bar\sigma})(1-N_{j,\sigma}-N_{j,\bar\sigma}+c_j)$,
$p_2=(1-N_{\ell,\bar\sigma})(N_{j,\bar\sigma}-c_j)$,
$p_3=(1-N_{\ell,\bar\sigma})(N_{j,\sigma}-c_j)$,
$p_4=(1-N_{\ell,\bar\sigma})c_j$,
$p_5=N_{\ell,\bar\sigma}(1-N_{j,\sigma}-N_{j,\bar\sigma}+c_j)$,
$p_6=N_{\ell,\bar\sigma}(N_{j,\bar\sigma}-c_j)$,
$p_7=N_{\ell,\bar\sigma}(N_{j,\sigma}-c_j)$, and
$p_8=N_{\ell,\bar\sigma}c_j$. The denominators for the eight
configurations are (i) $\Pi_1=\mu_{\ell}\mu_j-t^2_{c}$ with  both
dots empty, (ii) $\Pi_2=(\mu_{\ell}-U_{\ell,j})(\mu_j-U_j)-t^2_{c}$,
with dot A empty and dot B filled by one electron with spin
$\bar\sigma$, (iii)
$\Pi_3=(\mu_{\ell}-U_{\ell,j})(\mu_j-U_{j,\ell})-t^2_{c}$ with dot A
empty and dot B filled by one electron with spin $\sigma$, (iv)
$\Pi_4=(\mu_{\ell}-2U_{\ell,j})(\mu_j-U_j-U_{j,\ell})-t^2_{c}$ with
dot A is empty and dot B filled by two electrons, (v)
$\Pi_5=(\mu_{\ell}-U_{\ell})(\mu_j-U_{j,\ell})-t^2_{c}$ with dot B
empty and dot A filled by one electron with spin $\bar\sigma$, (vi)
$\Pi_6=(\mu_{\ell}-U_{\ell}-U_{\ell,j})(\mu_j-U_j-U_{j,\ell})-t^2_{c}$
with both dots filled by one electron with spin $\bar\sigma$,
$\Pi_7=(\mu_{\ell}-U_{\ell}-U_{\ell,j})(\mu_j-2U_{j,\ell})-t^2_{c}$
with dot A filled by one electron with spin $\bar\sigma$ and dot B
filled by one electron with spin $\sigma$, and (viii)
$\Pi_8=(\mu_{\ell}-U_{\ell}-2U_{\ell,j})(\mu_j-U_j-2U_{j,\ell})-t^2_{c}$
with dot A filled by one electron with spin $\bar\sigma$ and dot B
filled by two electrons.
$\mu_{\ell}=\epsilon-E_{\ell}+i\Gamma_{\ell}$. The notations
$E_{\ell}$, $U_{\ell}$, and $U_{\ell,j}$ denote, respectively, the
energy levels of dots, intradot Coulomb interactions, and interdot
Coulomb interactions. $t_c$ denotes the electron hopping strength
between two dots.

The probability factor for all channels of Eq.~(\ref{3}) are
determined by the thermally averaged one-particle occupation number and
two-particle correlation functions, which can be obtained by solving
the lesser Green's functions. We have $N_{\ell,\sigma}=-(1/\pi)\int
d\epsilon f_{\ell}(\epsilon)ImG^r_{\ell,\sigma}(\epsilon)$, and
$c_{\ell}=-(1/\pi)\int d\epsilon
f_{\ell}(\epsilon)ImG^r_{\ell,\ell}(\epsilon)$, where the retarded
Green functions $G^r_{\ell,\sigma}(\epsilon)$ and
$G^r_{\ell,\ell}(\epsilon)$ are, respectively, given by
\begin{eqnarray}
& &G^r_{\ell,\sigma}(\epsilon)\\ \nonumber &=& \frac{p_1}
{\mu_{\ell}-t^2_{c}/\mu_j }+\frac{p_2}
{(\mu_{\ell}-U_{\ell,j})-t^2_{c}/(\mu_j-U_j)}\\
\nonumber
&+&\frac{p_3}{(\mu_{\ell}-U_{\ell,j})-t^2_{c}/(\mu_j-U_{j,\ell})}\\
&+&\nonumber
\frac{p_4}{(\mu_{\ell}-2U_{\ell,j})-t^2_{c}/(\mu_j-U_j-U_{j,\ell})}\\
\nonumber
 &+& \frac{p_5}
{(\mu_{\ell}-U_{\ell})-t^2_{c}/(\mu_j-U_{j,\ell})}\\
\nonumber &+&\frac{p_6}
{(\mu_{\ell}-U_{\ell}-U_{\ell,j})-t^2_{c}/(\mu_j-U_j-U_{j,\ell})}\\
\nonumber
&+&\frac{p_7}{(\mu_{\ell}-U_{\ell}-U_{\ell,j})-t^2_{c}/(\mu_j-2U_{j,\ell})}\\
\nonumber
&+&\frac{p_8}{(\mu_{\ell}-U_{\ell}-2U_{\ell,j})-t^2_{c}/(\mu_j-U_j-2U_{j,\ell})},
\end{eqnarray}

and

\begin{eqnarray}
& & G^r_{\ell,\ell}(\epsilon)\\ \nonumber&=&
\frac{p_5} {(\mu_{\ell}-U_{\ell})-t^2_{c}/(\mu_j-U_{j,\ell})}\\
\nonumber&+&\frac{p_6}
{(\mu_{\ell}-U_{\ell}-U_{\ell,j})-t^2_{c}/(\mu_j-U_j-U_{j,\ell})}\\
\nonumber
&+&\frac{p_7}{(\mu_{\ell}-U_{\ell}-U_{\ell,j})-t^2_{c}/(\mu_j-2U_{j,\ell})}\\
\nonumber
&+&\frac{p_8}{(\mu_{\ell}-U_{\ell}-2U_{\ell,j})-t^2_{c}/(\mu_j-U_j-2U_{j,\ell})}.
\end{eqnarray}
Occupation numbers of Eqs. (3) and (4) should be solved
self-consistently. Note that in the absence of $t_c$ the expressions
of Eqs. (3) and (4) can also be found in our previous
works.$^{16,17}$ In the linear response regime, Eq. (2) can be
rewritten as
\begin{equation}
J={\cal L}_{11} \Delta V+{\cal L}_{12} \Delta T,
\end{equation}
where $\Delta T=T_L-T_R$ is the temperature difference across the
junction. Coefficients in Eq. (5) are given by
\begin{eqnarray}
{\cal L}_{11}&=&\frac{2e^2}{h} \int d\epsilon {\cal T}(\epsilon)
(\frac{\partial f(\epsilon)}{\partial E_F})_T,\\ \nonumber {\cal
L}_{12}&=&\frac{2e}{h} \int d\epsilon {\cal T}(\epsilon)
(\frac{\partial f(\epsilon)}{\partial T})_{E_F}.
\end{eqnarray}

Here ${\cal T}(\epsilon)$ and
$f(\epsilon)=1/[e^{(\epsilon-E_F)/k_BT}+1]$ are evaluated at thermal equilibrium. If the system is in an
open circuit, the electrochemical potential ($\Delta V$) will be established
in response to a temperature gradient; this electrochemical
potential is known as the Seebeck voltage. The Seebeck coefficient
is defined as $S=\Delta V/\Delta T=-{\cal L }_{12}/{\cal L}_{11}$,
where ${\cal L}_{11}$ denotes the electrical conductance, $G_e$.

Using the following physical parameters: $U_{\ell}=U_0=30\Gamma_0$,
$U_{12}=10\Gamma_0$, and $\Gamma_L=\Gamma_R=1\Gamma_0$, where the
average tunneling rate $\Gamma_0$ has been used as a convenient
energy unit, we numerically calculate the thermoelectric
coefficients ${\cal L}_{11}$ and ${\cal L}_{12}$. Figure 1 shows the
electrical conductance $G_e$ and Seebeck coefficient (S) as a
function of temperature at nonzero orbital offset ($E_1-E_2=\Delta E
\neq 0$). For the case of $E_2=E_F-U_0$ and $\Delta E=U_2-U_{12}$,
the electrons injected by a small bias can only tunnel through the
the spin singlet state of the SCQD. This is the so-called "spin
blockade" effect of the SCQD.$^{14,15,18,19)}$ The electrical
conductance is suppressed when $E_2$ deviates from the resonance
condition with $\Delta E=U_2-U_{12}$, and we observe that there is a
zero-crossing temperature $T_0$ for the Seebeck coefficient, i.e,
$S(T)=0$ at $T=T_0$. A zero Seebeck coefficient $S(T_0)$ indicates
that the current arising from the temperature gradient can be
self-consistently balanced without electrochemical potential. The
negative S indicates that electron carriers of the left (hot)
electrode diffuse into the right (cold) electrode via the resonant
channels above $E_F$, the negative $\Delta V $ is built up to reach
the condition of $J=0$ at open circuit [see eq. (5)]. For example,
the curve of $E_1=E_F-10\Gamma_0$ and $E_2=E_F-30\Gamma_0$  has a
negative Seebeck coefficient. Note that the resonant channel
$E_1+U_{12}=E_2+U_2$ [see the inset of Fig. 1(b)] has a zero
contribution in S. On the other hand, the Seebeck coefficient is
positive when holes of the right electrode diffuse into the left
electrode via the resonant channels below $E_F$. Here, we define the
unoccupied states below $E_F$ as holes. Consequently, the change in
sign of S is attributed to the competition between tunneling
currents due to electrons and holes. To further clarify the
mechanism of S in sign change with respect to temperature, we
consider the case of zero orbital offset for simplicity.

Figure 2(a) shows the Seebeck coefficient as a function of
temperature for various electron Coulomb interactions with
$E_{\ell}=E_0=E_F-10\Gamma_0$. In the noninteracting case
($U_{\ell}=U_{\ell,j}=0$), S is always positive (see dash-dotted
curve). This is because the tunneling process is dominated by holes
of the hot electrode (left electrode) diffusing into the cold
electrode (right electrode) through a level $E_0$ below $E_F$ in the
weak $t_c$ limit. To reveal the mechanism of sign change in the
Seebeck coefficient for the interacting case (as shown by solid
curves in Fig. 2), we analyze the contributions associated with
different poles described in eq. (2). In the weak $t_c$ limit, there
are six poles associated with the left dot for the spectral density:
$\epsilon=E_0$, $\epsilon=E_0+U_{12}$, $\epsilon=E_0+2U_{12}$,
$\epsilon=E_0+U_0$, $\epsilon=E_0+U_0+U_{12}$, and
$\epsilon=E_0+U_0+2U_{12}$. In addition, we also find that only four
channels ($p_1$, $p_3$, $p_6$, and $p_8$) have high probability
weighting. They correspond, respectively, to the resonant channels
$\epsilon=E_0$, $\epsilon=E_0+U_{12}$, $\epsilon=E_0+U_0+U_{12}$,
and $\epsilon=E_0+U_0+2U_{12}$. When $U_{12}=10\Gamma_0$, the $p_3$
channel does not contribute to S because its pole location is
aligned with the Fermi level, i.e, $E_0+U_{12}=E_F$. The strengths
of $p_1$, $p_6$, and $p_8$ for $U_{12}=10\Gamma_0$ as functions of
temperature are shown in Fig. 2(b). The hole contribution is given
by $p_1$, which leads to a positive contribution to $S$. The
electron contribution is governed by $p_6$ and $p_8$, which
correspond to resonant channels with energy above $E_F$, providing a
negative contribution to the Seebeck coefficient. Consequently, the
change in sign of $S$ results from the interplay between the
competition of electron and hole flows. The behavior of the Seebeck
coefficient near the zero-crossing temperature $T_0$ is linear. When
the equilibrium temperature is away from $T_0$, the sign change in
the Seebeck coefficient indicates that one can produce
thermoelectric current in either the forward or reverse direction,
leading to a bipolar thermoelectric effect. On the basis of the
closed form solutions of transmission factors and Green's functions,
we can solve ${\cal L}_{11}$ and ${\cal L}_{12}$ in terms of
polygamma functions to find $T_0$ accurately. We find that $T_0$ is
mainly dominated by $U_{\ell}$ in the atomic limit ($t_c/U_{\ell}\ll
1$), and $T_0$ increases with increasing $U_{\ell}$ at fixed
$E_{\ell}$, but is insensitive to the variation in $U_{12}$. These
results imply that the SCQD may have a stable $T_0$ under small
fluctuations of the QD size, and a high value of $T_0$ can be
achieved for small size QDs .

To achieve a thermal device with a high density of charge and heat
currents, we need to consider a higher SCQD density. Consequently,
the proximity effect between SCQDs on the Seebeck coefficient should
be investigated. For simplicity, we employ the case shown in the
inset of Fig. 3 as an example. The detailed expression of
transmission factor for two parallel SCQDs can be found in ref. 15,
where we investigated the charge ratchet effect on current
rectification. Figure 3 shows the Seebeck coefficient as a function
of temperature. The curves $S_1(T)$, $S_2(T)$, $S_3(T)$, and
$S_4(T)$ correspond respectively to the interdot Coulomb interaction
$U=0,4,6,$ and $8\Gamma_0$. In Fig. 3(a), we consider
$U_{\ell}=30\Gamma_0$, $E_{\ell}=E_F-10\Gamma_0$, and
$U_{12}=U_{34}=10\Gamma_0$. In Fig. 3(b), we consider
$U_{\ell}=60\Gamma_0$, $E_{\ell}=E_F-20\Gamma_0$, and
$U_{12}=U_{34}=20\Gamma_0$. We observe that $T_0$ is pushed toward a
lower temperature with increasing U. In Fig. 3(a), $S_4(T)$ (blue
line) is negative in the entire temperature regime. Such a behavior
indicates that the number of resonant channels involving electron
Coulomb interactions above $E_F$ increases with increasing U and
they dominate electron carrier transport. In Fig. 3(b), we obtain
$k_BT_0=27\Gamma_0$ in the curve of $S_1(T)$. Compared with the
black line in Fig. 3(a), $T_0$ is enhanced. This is attributed to
the increase in intradot Coulomb interactions $U_{\ell}$. In
Particular, there are two zero-crossing temperatures in the $S_2(T)$
and $S_3(T)$ curves. They are $k_BT_0=6\Gamma_0$ and
$k_BT_0=22\Gamma_0$ in the $S_2(T)$ curve, and $k_BT_0=5\Gamma_0$
and $k_BT_0=19\Gamma_0$ in the $S_3(T)$ curve. The oscillatory
behavior of the Seebeck coefficient in the $S_2(T)$ and $S_3(T)$
curves is observed. There is only one zero-crossing temperature,
$k_BT_0=14.5\Gamma_0$, in the $S_4(T)$ curve.

In this study, we find that the sign of electrochemical potential
can be tuned by selecting the equilibrium temperature for a given
temperature gradient. This implies that a temperature-controlled
bipolar thermoelectric device can be achieved. For the two parallel
SCQDs, the oscillatory behavior of the Seebeck coefficient with
respect to temperature is also observed.

{\bf Acknowledgments}\\
This work was supported in part by the National Science Council of
the Republic of China under Contract Nos. NSC 99-2112-M-008-018-MY2,
and NSC 98-2112-M-001-022-MY3.

\mbox{}\\
${}^{\dagger}$ E-mail address: mtkuo@ee.ncu.edu.tw\\
$^*$ E-mail address: yiachang@gate.sinica.edu.tw

\mbox{}

\newpage

{\bf Figure Captions}

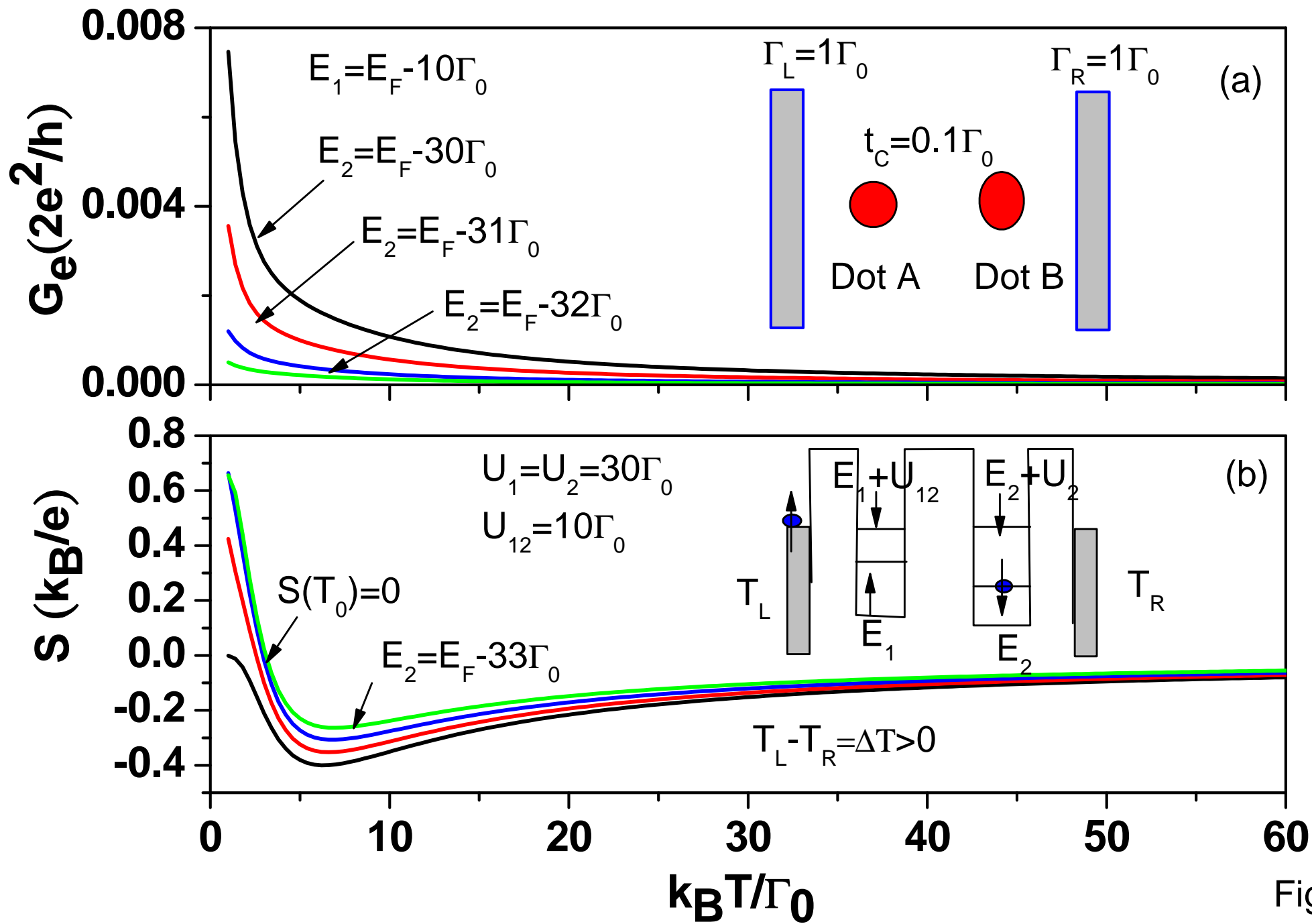
Fig. 1. The electrical conductance $G_e$ and Seebeck coefficient as
a function of temperature for various values of $E_2$ (from
$E_F-10\Gamma_0$ to $E_F-33\Gamma_0$) with $E_1=E_F-10\Gamma_0$,
$t_C=0.1\Gamma_0$ and $\Gamma_L=\Gamma_R=1\Gamma_0$. Insets shown in
Fig. 1(a) and 1(b) illustrate, respectively, the SQCD connected to
the metallic electrodes and the band diagram corresponding to Fig.
1(a).

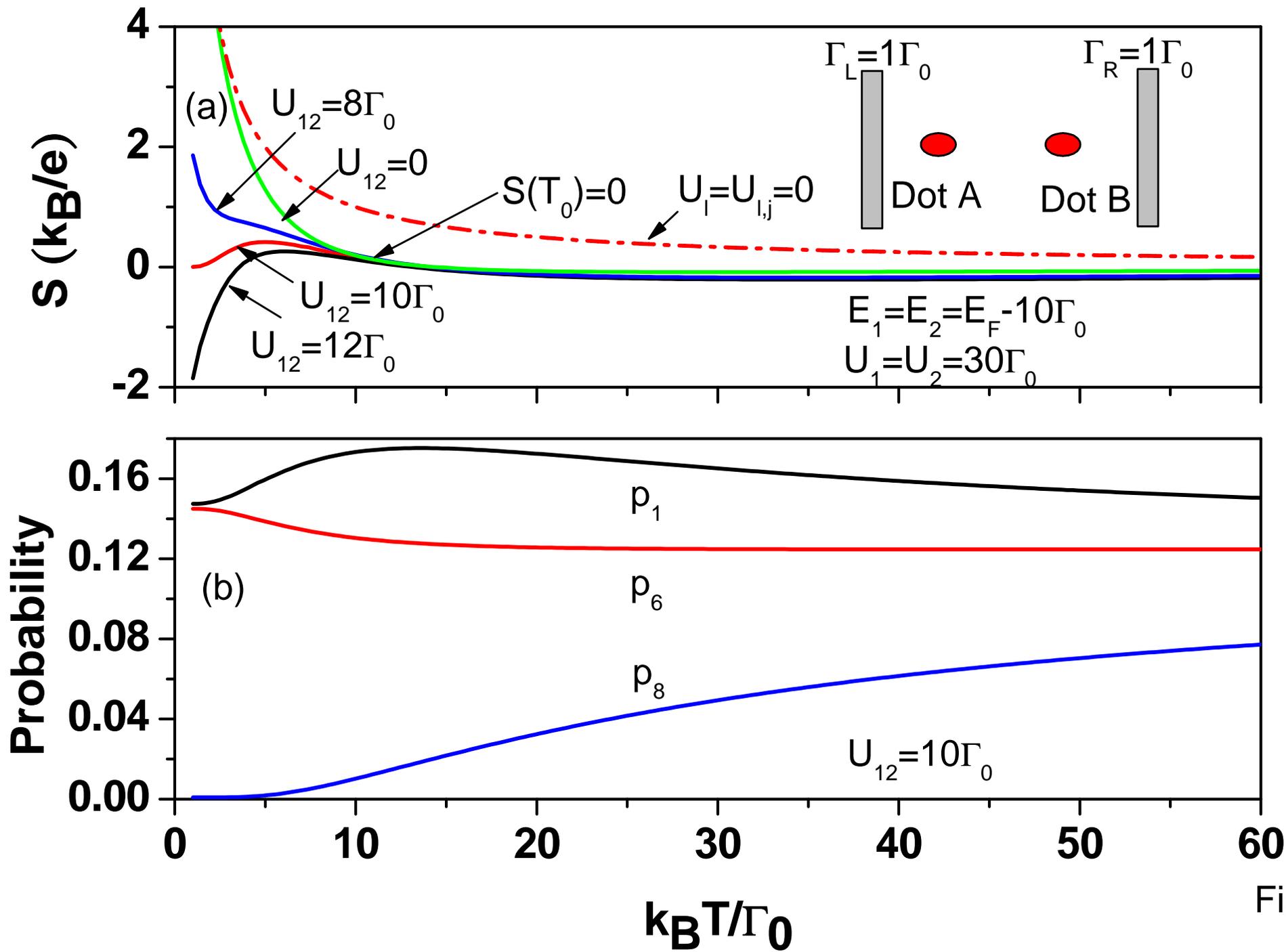
Fig. 2. Seebeck coefficient as a function of temperature at
$E_1=E_2=E_F-10\Gamma_0$ for different electron Coulomb
interactions. Other parameters are the same as those of Fig. 1.
Diagram (b) shows the probability of resonant channels in the case
of $U_{12}=10\Gamma_0$.

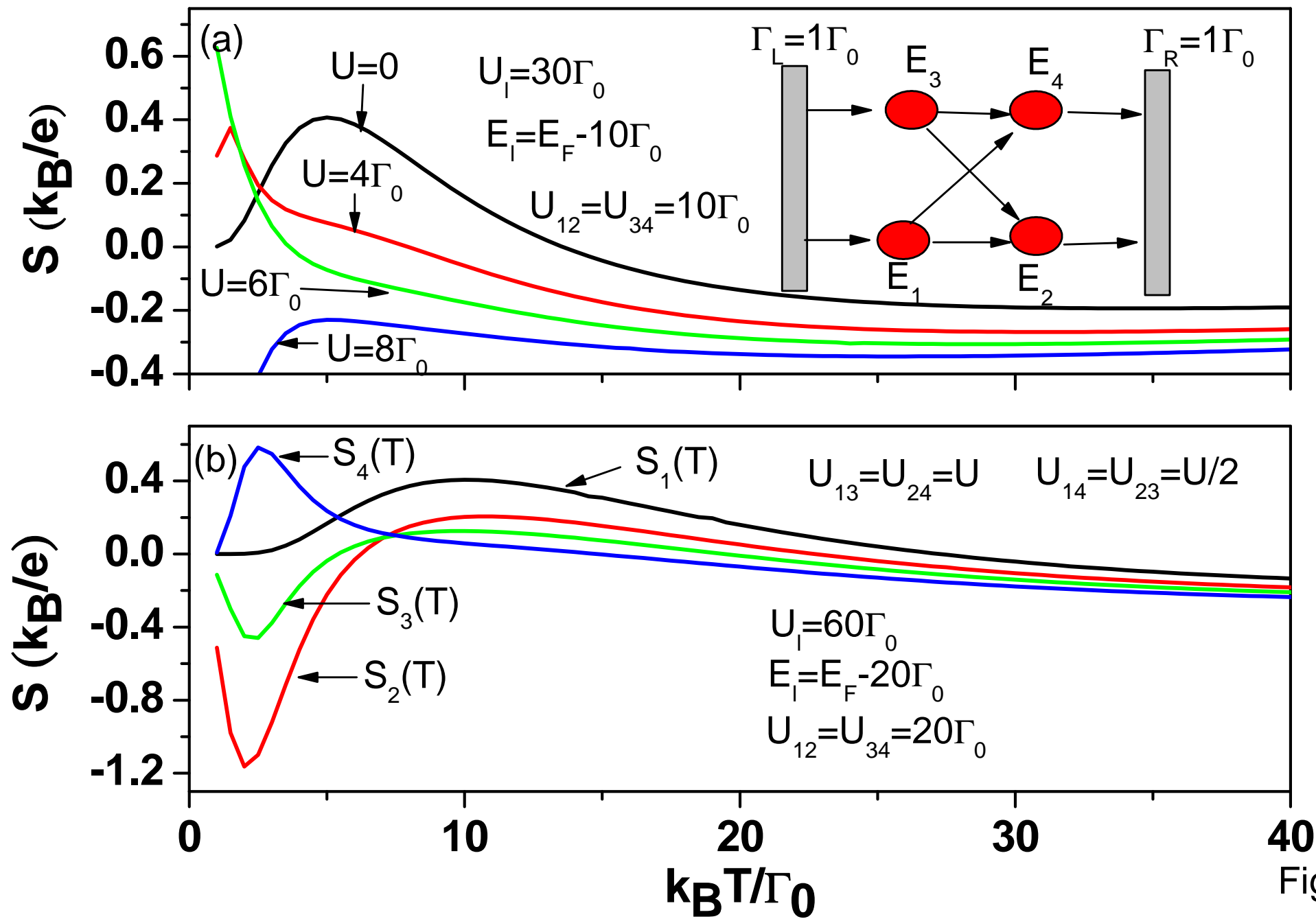
Fig. 3.  Seebeck coefficient as functions of temperature for various
values of interdot Coulomb interaction $U_{13}=U_{24}=U$ and
$U_{14}=U_{24}=U/2$. Diagrams (a) and (b) consider two different
sets of physical parameters $U_{\ell}=30\Gamma_0$,
$E_{\ell}=E_F-10\Gamma_0$, $U_{12}=U_{34}=10\Gamma_0$ and
$U_{\ell}=60\Gamma_0$, $E_{\ell}=E_F-20\Gamma_0$,
$U_{12}=U_{34}=20\Gamma_0$, respectively.

\end{document}